\documentclass[]{amsart}

\usepackage{wrapfig}
\usepackage{amsmath,amssymb, amsfonts, mathtools}
\usepackage{amsthm}

\usepackage{newtxtext,newtxmath}       
\usepackage{helvet}         
\usepackage{courier}        
\usepackage{graphicx}    
\usepackage{subcaption}
\usepackage{multicol}        
\usepackage[bottom]{footmisc}  
\usepackage{amsfonts}       
\usepackage{verbatim, booktabs}
\usepackage{bbm, bm}
\usepackage{multirow}
\usepackage{csvsimple}
\usepackage{longtable}
\usepackage[noabbrev,capitalise]{cleveref}
\usepackage{algpseudocode, algorithm}

\usepackage{amsmath,amssymb,verbatim,amsthm,mathtools}

\usepackage{subcaption}
\usepackage{enumitem}
\usepackage{epstopdf}
\usepackage{varwidth}
\usepackage{algorithm}
\usepackage{color}
\usepackage[]{xcolor}

\graphicspath{ {./images} }

\newcommand{\R}{\mathbb{R}}

\newcommand{\given}{\,|\,}
\newcommand{\der}{\mathrm{d}}

\ifpdf
\DeclareGraphicsExtensions{.eps,.pdf,.png,.jpg}
\else
\DeclareGraphicsExtensions{.eps}
\fi

\begin{document}
\title{Data Assimilation for Robust UQ Within Agent-Based Simulation on HPC Systems}

\author{Adam Spannaus$^1$}
    \email[Corresponding author]{spannausat@ornl.gov}
\thanks{This manuscript has been authored by UT-Battelle, LLC under Contract No. DE-AC05-00OR22725 
	with the U.S. Department of Energy. The United States Government retains and the publisher, 
	by accepting the article for publication, acknowledges that the United States Government 
	retains a non-exclusive, paid-up, irrevocable,world-wide license to publish or reproduce 
	the published form of this manuscript, or allow others to do so, for United States 
	Government purposes. The Department of Energy will provide public access to these 
	results of federally sponsored research in accordance with the DOE Public 
	Access Plan (http://energy.gov/downloads/doe-public-access-plan).}
\address[1]{Oak Ridge National Laboratory, Oak Ridge, TN 37830}
\author{Sifat Afroj Moon$^1$}
\author{John Gounley$^1$}
\author{Heidi A. Hanson$^1$}

\renewcommand{\shortauthors}{Spannaus et al.}


\begin{abstract} 

    Agent-based simulation provides a powerful tool for 
    \textit{in silico} system modeling. However, these simulations do not provide built-in
    methods for uncertainty quantification (UQ). 
    Within these types of models
    a typical approach to UQ is to 
    run multiple realizations of the model then compute aggregate statistics. This approach is limited due to
    the compute time required for a solution. When faced with an 
    emerging biothreat, public health decisions need to be made quickly and 
    solutions for integrating near real-time data with analytic tools are needed.

    We propose an integrated Bayesian UQ framework for agent-based models based on sequential 
    Monte Carlo sampling. Given streaming or static data about 
    the evolution of an emerging pathogen,
    this Bayesian framework provides a distribution over the parameters 
    governing the spread of a disease through a population. These  
    estimates of the spread of a disease may be provided to public health agencies seeking to abate the spread.  

    By coupling agent-based simulations with Bayesian modeling in a data assimilation, our proposed 
    framework provides a powerful tool for
    modeling dynamical systems \textit{in silico}.
We propose a method which reduces model error and provides a range of realistic possible outcomes. 
Moreover, our method addresses two primary limitations of ABMs: the lack of UQ and an
inability to assimilate data.
    Our proposed framework combines the flexibility of an agent-based
    model with UQ provided by the Bayesian paradigm in a workflow which scales well to HPC systems. We 
    provide algorithmic details and results on a simulated outbreak
    with both static and streaming data.

\end{abstract}


\maketitle

\section{Introduction}
\label{sec:intro}
Infectious disease modeling 
creates simulations of ongoing or
potential disease outbreaks. Assuring accurate estimates for public health agencies seeking input 
and guidance on the future course of disease is critical during an outbreak. Agent-based modeling (ABM) is a commonly used tool that allows for realistic simulations of 
disease spread, hospital over-capacity, and mortality when different public health 
intervention strategies are implemented. While a effective 
modeling tool, ABMs do not have built-in uncertainty 
quantification (UQ). The traditional solution is an ensembling approach, which requires running a large number of instances of each simulation and can have large computational costs and delay the time to solution.  
For example, an agent-based system with millions of heterogeneous agents 
requires a significant amount of computational time for just one realization, even in a high-performance computing (HPC) setting~\cite{bhattacharya2024novel}.

While running multiple simulations to aggregate statistics is a trivially parallel operation, it cannot quantify
the parameter estimates of the rates controlling movement between states in an epidemiological model, which are typically specified \emph{a priori}. 
As the model steps forward in time, it can reflect the range of outcomes 
conditioned on the parameters governing the ABM simulation itself. Given a shock to the 
real-world system,
such as the emergence of a more virulent strain, the ABM results would fail to reflect this
new reality. To sufficiently quantify the new strain's impact on a population
the ABM simulation would require a new set of simulations with different parameters.
Moreover, the parameters controlling the movement through the different compartments in an 
epidemiological model may change with time. For example, treatment regimes may improve as medical professionals have time to study and understand a novel disease, or vaccines may 
be introduced, reducing the transmissibility of a disease within a proportion of the population.
Assimilating real-world data in an online or near-real time fashion would help mitigate some
of the shortcomings of ABMs, but these types of models do not have any mechanism to smoothly
integrate data, nor a mechanism to update their parameters in response to environmental changes. 

The Bayesian paradigm offers solutions to the shortcomings of traditional ABM approaches.
By providing distributional information about the 
disease parameters and the proportion of a population in each of the epidemiological 
model's states at each time step, Bayesian methods
are a natural way to seamlessly integrate data into a simulation.
In particular, sequential Monte Carlo (SMC) methods 
allow for Bayesian inference within a time series and are widely used in signal tracking,
numerical weather prediction, and geophysical applications. This
class of methods are widely
applicable in inference problems arising in 
hidden Markov models with possibly noisy and/or partially hidden observations. 

\begin{figure}
        \centering
        \includegraphics[width=1\linewidth] {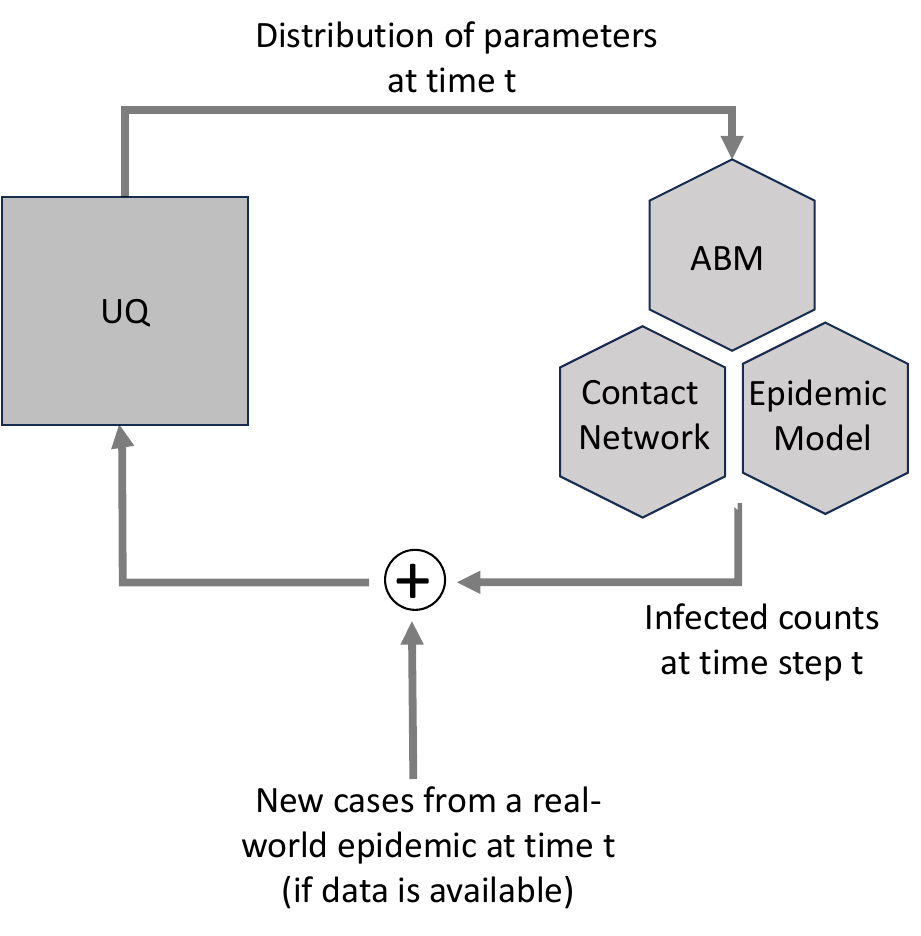}
        \caption{Illustration of the coupled ABM and UQ framework. Information
        is passed between the different methods, yielding high-fidelity ABM
        simulations and UQ of the model parameters.}
        \label{fig:framework}
    \end{figure}

There are two limitations with SMC methods which have previously restricted their use. 
First, this class of methods require a `resampling' step to ensure they capture of 
the full range of possibilities.
This step ensures that the posterior distribution does not degenerate to a point mass and
fails to explore the sample space~\cite{del2006sequential}.  
Moreover, resampling presents a computational bottleneck as it requires a synchronization and reduction step
across all processes; the works of~\cite{murray2016parallel,whiteley2016role} seek to lessen 
this computational burden. 
Second, designing bespoke proposal distributions allowing for optimal inference in 
high-dimensional space. In this research, we couple a parallel implementation with limited
communication for the resampling scheme to join the individual level
fidelity of an ABM with the Bayesian UQ,  
and defer discussion of the latter
to a future work. Figure~\ref{fig:framework} presents an illustration of our coupling framework.

Our proposed method reduces model error and provides a range of realistic possible outcomes. 
This work addresses two limitations of ABMs: 1.) 
an inability to assimilate streaming data into the simulation, accounting for a more virulent strain emerging,
improved medical treatments, or inaccessibility of care due to hospital over-capacity; and 2.)
the lack of scalable UQ methods. Our method allows for UQ
and periodic correction within an ABM, by assimilating real-world data into the solution.
Consequently, our proposed method yields results closer to reality as a situation evolves. 
The codebase is a SMC implementation in Jax, providing 
  an HPC-ready Python package, based in part on~\cite{particles2020github}. 
  Existing C++ template 
  libraries~\cite{zhou2015vsmc,johansen2009smctc,zhou2018mckl}
do not provide a parallel-prefix sum, nor is it available within 
  Python implementations~\cite{PyMC,particles2020github}. Additionally,
  the flexibility of Jax provides performance portability with
  one codebase on CPUs and GPUs.

\subsection{Related Works}
\label{ssec:related}
Other methods have been proposed for incorporating ABMs within a UQ 
framework, see~\cite{asher2023dynamic,cocucci2022inference,fadikar2024improveduncertaintyquantificationstochastic} for example.
These proposed solutions are often not effective in a real world scenarios however,
due to Gaussian assumptions on the structure of the problem, 
linearity assumptions on the system dynamics, or
numerical approximations within the UQ method itself.

The works of Cocucci et al. and Asher et al. propose two 
methods to 
to calibrate an ABMs parameters,
either through some assumptions on the uncertainties present within the 
epidemiological model or an approximate Bayesian computation (ABC) 
framework. 
Asher et al.~\cite{asher2023dynamic} describe a (near) real-time dynamic ABM calibration
obtained via an ABC setting.
Their methodology 
is applied to a large-scale epidemiological ABM, and had previously only 
been applied to small toy problems.

Cocucci et al.~\cite{cocucci2022inference} combine an epidemiological ABM
with an ensemble Kalman filter in a data assimilation framework. The Gaussian
assumption underlying the Kalman filter works well if the uncertainties are
normally distributed, but the method may fail to capture the disease dynamics 
in the early or late stages of an 
outbreak when there are low rates of infection~\cite{ebeigbe2020poisson}. 
Moreover, the 
Monte Carlo approximations of the distributions of interest may fail to converge
asymptotically in the case of non-linear dynamics~\cite{law2015data},
which is the setting of an epidemiological model.

Fadikar et al. present a method that is similar to ours~\cite{fadikar2024towards}.
They employ a importance resampling strategy, 
but do not address the computational bottlenecks 
inherent to importance resampling methods. Our previous 
work~\cite{spannaus2022inferring} and that of Del Moral et al.~\cite{del2015sequential} 
use SMC coupled with the
Particle marginal Markov Chain Monte Carlo methodology~\cite{andrieu2010particle}
to simultaneously infer the epidemiological time-series and governing parameters
of the epidemiological model. The primary limitation of these methods is the inability to easily assimilate real-time streaming 
data into the Markov chain Monte Carlo sampling scheme.

Lastly, the issue of parallel resampling within SMC has been previously 
addressed by Murray et al.~\cite{murray2016parallel},
who proposed a few different methods, each 
with its own set of advantages and limitations. The authors of~\cite{guldas2015practical}
describe a divide-and-conquer algorithm for efficient resampling within 
a SMC framework. Our present work
uses a CUDA-accelerated implementation of the prefix sum algorithm by Blelloch et al.~\cite{blelloch1990prefix}
to compute the cumulative sum over the vector in the 
high-dimensional weight space in the SMC sampling method.

\subsection{Contributions}

    We describe our approach to provide ABM with scalable Bayesian data assimilation and UQ.
        Within the data assimilation framework, we connect an ABM model with a Bayesian formulation, integrating data with mathematical models. Data
    assimilation is a well-established paradigm within the atmospheric and oceanographic modeling communities~\cite{law2015data}, and
    gives equal weight to both data and theoretical models. 
    Existing SMC libraries in both Python and C++ do not 
    provide a parallel prefix 
    sum, nor are they easily ported to GPU implementations~\cite{zhou2015vsmc,johansen2009smctc,zhou2018mckl,PyMC,particles2020github}. Our Jax codebase coupled with the Thrust library
    provides flexibility for
    one codebase to run on either a CPU or GPU~\cite{Thrust}.
    
    In the proof-of-concept experiments considered here,
    we do not have empirical data. 
    Instead we use a small collection of ABM realizations over a short time window, i.e., a few days to one week, which are 
    used as input to a SMC sampler over the same time span. This sampling method yields  
    distributional estimates of the population proportion in each bin of a
    compartmental epidemiological model and the parameters governing the rate of 
    movement between those bins. 
    Lastly, any ABM simulation requires the choice of an epidemiological model \textit{a prioiri}. When faced with an emerging
    outbreak, sufficient data may not exist to choose one model over another; by framing the problem within the 
    Bayesian framework, model selection based upon the available data is 
    straightforward. One incurs only the 
    additional computational cost of running different simulations in parallel~\cite{kass1995bayes}.

    The outline of the paper is as follows. In~\cref{sec:methods} we give a 
    high-level description of the ABM and SMC methods used. Our algorithmic design is
    presented in~\cref{sec:the_algorithm} and numerical results are in~\cref{sec:numerics}.
    We conclude with discussion in~\cref{sec:discussion}.

\section{Methods}
\label{sec:methods}

\subsection{Epidemiological Modeling}

One of the most widely 
used epidemiological models of disease transmission is the Susceptible-Infected-Removed 
(SIR) compartmental model~\cite{kermack1927contribution}. This model describes the 
movement of a population between three compartments: those who are
susceptible to an infection, those who are actively infected, and 
those who have either recovered from the disease or are deceased due to it.
The SIR model has been augmented to include other compartments,
such as an exposed individuals, those who have been 
exposed but are not actively infected, 
asymptomatic carriers, or
individuals immune to the infection~\cite{blackwood2018introduction}.
An ordinary differential equation model of an
epidemiological system model assumes population homogeneity and consequently allows
for only a small number of variables to describe the system. It tracks the 
number of individuals in each compartment at each time step. 
These macro-state variables smooth out the noise in the dynamical system when
viewed at the individual level, as through an ABM.
The individual-level nature of the ABM  drives the stochasticity present within the 
simulation. 

\subsubsection{Individual-Based Network Epidemic Modeling}
\label{ssec:abm}

Agent-based modeling is a systematic way to represent heterogeneous, interactive complex 
systems~\cite{eubank2004modelling,toroczckai2005agent,barrett2008episimdemics}. It is a 
useful decision tool in the public health domain to simulate individual-level disease dynamics and agents behaviour. Previous literature has used 
agent-based simulation to model numerous diseases and complex interventions, such as 
targeted vaccination, contact tracing, and quarantine strategies~\cite{espinoza2023coupled,afroj2023all}. 

In this 
work, agents are individual people and connections between them are represented by a social 
contact network, denoted $G (V, E)$~\cite{seshadhri2012community}. 
We define $V$ as the set of agents and $E$ as the set of 
undirected edges or links between two agents.
A node $v_i\in V$ in $G$ represents an agent. A direct contact 
between two agents is represented by an edge $e_i\in E$. For the SIR model, agents are divided into 
each of the three states: susceptible (S), infected (I), and recovered (R). 
A susceptible agent can contract the disease from an infected agent through direct contact. 
The transition of an agent from the susceptible to the infected state ($S \rightarrow I$) 
depends on the health states of its neighboring agents. In contrast, the transition of an 
agent from the infected to the recovered state ($I \rightarrow R$) does not depend on its 
neighbors. $H(t)$ keeps track of health state of all agents at time $t$. The dimension of 
of $H(t)$ is $|V| \times 1$ where $|V|$ is the number of agents and an element of $H(t)$ 
is denoted 
$H(t)[v_i]\in\{S,I, R\}$. As a preprocessing step, we divide the agents into multiple MPI (Message Passing Interface) processes. Algorithm~\ref{alg:abm} outlines the agent-based simulation 
process for the spread of an SIR model at time $t$ in a single parallel MPI process. After each time step, we synchronize the health state of all agents across all processes.

\begin{algorithm}
        \caption{Agent-based simulation step at time $t$ for the individual-based network SIR epidemic model in a parallel process.}\label{alg:abm}.
        \begin{algorithmic}[1]
        
         \Statex \textbf{Input: } Network $G(V,E)$;
         \Statex \hspace{3.2em} Set of susceptible nodes at time $t$ in the current process,\Statex  
         \hspace{3.2em}$V_S(t)$; 
         \Statex \hspace{3.2em}Set of infected nodes at time $t$ in the current process, \Statex  
         \hspace{3.2em}$V_I(t)$;
         \Statex \hspace{3.2em}Set of recovered nodes at time $t$ in the current process, \Statex  
         \hspace{3.2em}$V_R(t)$;
         \Statex \hspace{3.2em}Health state of nodes at time $t, H(t)$;
         \Statex \hspace{3.2em}Distribution of disease model parameters, $\beta(t)$ and $\gamma(t)$ \Statex \hspace{3.2em}for SIR compartmental model 
         at time t, and S, I, R compartments.
        
            \State New infected nodes $N_I\gets[]$ 
            \State New recovered nodes $N_R\gets[]$ 
            
            \For{$ v \in V_I(t)$}
                \For {$nbr\in neighbors\_of\_v$ in G}
                    \If {$H(t)[nbr]==S$  $\&$ $nbr$ is in the same process of $v$}
                        \State $r_1 \gets \mathcal{U}[0,1]$ // generate a random number
                        \State $r_\beta \gets $ generate a number from the distribution of $\beta(t)$
                        \If {$r_1\leq \frac{r_\beta}{avgD(G)}$} // $avgD(G)$ is the average degree of the contact network $G$ 
                            \State $H(t)[nbr] \gets I$
                            \State $N_I \gets N_I+nbr$
                            \State $V_S(t) \gets V_S(t)-nbr$
                            
                        \EndIf
                        
                    \EndIf
                \EndFor
                \State $r_2 \gets \mathcal{U}[0,1]$ // generate a random number
                \State $r_\gamma \gets $ generate a number from the distribution of $\gamma(t)$
                \If{$r_2 \leq r_\gamma$}
                \State  $H(t)[v] \gets R$
                \State $N_R \gets N_R+v$
                \State $V_I(t) \gets V_I(t)-v$
                \EndIf
               
        \EndFor
        \State $V_I(t) \gets V_I(t)-N_I$
        \State $V_R(t) \gets V_R(t)-N_R$\\
        \Return $H(t)$, $V_S(t)$, $V_I(t)$, and $V_R(t)$

        \end{algorithmic}
    \end{algorithm}

\subsubsection{The Need for Data Assimilation}

In ABM, models are defined at the micro scale as a set
of values assigned to each agent at some time $t$. The model defines rules for how these agents interact and how their values evolve with time.
For public-health agencies, the specifics of these interactions are not
as meaningful as the macro-scale results, e.g., summary
statistics of the number of infected individuals 
or aggregated quantities related to the progression of a disease, such as the 
percent of hospital capacity in use.
While these quantities represent the population as a whole 
at the macro-level, the goal of the simulation is to faithfully 
represent the population at the micro-level, but the observations,
either from streaming real world data or \textit{in silico}
experiments, 
are not seamlessly integrated into the individual-level simulation. 
This poses the issue of identifiability,
where a bijective mapping between the macro and individual states
does not exist. 
Consequently, we integrate observations into the macro-scale simulations
and perform the data assimilation at this level, and will 
pass the resulting distributional estimates back into
the individual-based models, see~\cref{fig:framework}.
SMC periodically resamples
the ensemble of particles, which are in the epidemiological 
model's state space, bypassing the issue of identifiability  
and the need to adjust the epidemiological compartments in the ABM population 
at each data assimilation step~\cite{cocucci2022inference}.
Additionally, we chose to discretize the time steps within the SMC simulation 
at a more granular level than at the ABM simulation, to better identify changing 
disease trends.

\subsection{Data Assimilation}\label{ssec:smc}

    Data assimilation is a methodology for probabilistic inference within 
    noisy dynamical systems with partial observations. One of the primary
    tools for inference in this setting is SMC.
    Sequential Monte Carlo is a sampling technique used to sequentially analyze 
    a sequence of distributions of a state-space model. Informally, a 
    state-space model relates two discrete or continuous time processes,
    known as the hidden and observation processes,
    through a probabilistic model which incorporates densities for 
    both the hidden 
    and observation processes~\cite{doucet2001sequential}. Traditional sampling methods,
    such as Markov chain Monte Carlo, struggle with sampling from these high-dimensional 
    distributions due to correlations between time steps and identifying efficient ways
    to explore the sample space. SMC methods iteratively generate samples from a 
    sequence of distributions, such as those defined through a dynamical system. These methods
    are frequently augmented with a data assimilation step, where
    potentially noisy or sparse observations are incorporated into the algorithm, providing external feedback
    to the method about some quantities of interest.

    Formally, we consider a continuous time Markov process $X_t\in\mathcal{X}\subset\R^d$. 
    For a sequence of times $0=t_0,\dots,t_n=T$ we will write
    $y_{1:t} = \{y_\tau\}_{\tau=1}^t$ for a sequence $\{t_i\}_{i=1}^k$
    with $k\leq n$ and drop the time subscript for clarity of notation.
    At time $t$ we are interested in inferring information about
    some quantity $x_t$ through a possibly indirect observation $y_t$, which may be 
    corrupted by noise and/or be
    partially observed. Mathematically this takes the form:
    \begin{align}\label{eq:hmm}
        X_{t} &\sim p(x_{t}\given x_{t-1}, \theta_x) \\
        Y_{t} &\sim p(y_t\given x_{t}, \theta_y)
    \end{align}
    for $t\geq 1$, 
    assuming  $X_t\given X_{t-1}$ and $Y_t\given X_t$ admit densities $f(x_t\given x_{t-1}; \theta_x)$
    and $g(y_t \given x_t; \theta_y)$ respectively, for
    parameters $\theta_x, \theta_y$ and $X_0\sim\mu(x_0)$,
    the prior distribution of $x_0$, denoted by $\mu$. 
    Now~\cref{eq:hmm} with the prior $\mu$ defines a Bayesian model over the hidden Markov process $\{X_t\}_{t=0}^T$
    by considering the posterior parameterized by $\theta:=(\theta_x, \theta_y)$ written
    \begin{equation}\label{eq:post}
    	p_\theta(x_{0:T}\given y_{1:T}) = \frac{p_\theta(x_{0:T}, y_{1:T})}{p_\theta(y_{1:T})},
    \end{equation}
    which follows from the standard decompositions
   	\begin{align*}
		p_\theta(x_{0:T}, y_{1:T}) =& \mu(x_0)\prod_{t=1}^T\,g(y_t \given x_t; \theta_y)\prod_{t=1}^T f(x_t\given x_{t-1}; \theta_x)\\
		\intertext{and}
		p_\theta(y_{1:T}) =& \prod_{t=1}^T\,p(y_t\given y_{0:t-1};\theta_y)\\
		=& \int_{\mathcal{X}^{T+1}}\,\mu(x_0)\prod_{t=1}^T\,g(y_t \given x_t; \theta_y)\prod_{t=1}^T f(x_t\given x_{t-1}; \theta_x)\der x_{0:T}.
	\end{align*}

    If we consider the case of indirect observations, i.e., $Y_t\sim p(y_t\given x_t;\theta_y)$
    then the goal is to sequentially sample
    the posterior distribution of the hidden states, i.e.,  $\{ p(x_{0:t}\given Y_{1:t}=y_{1:t})\}_{t=1}^T$ in~\cref{eq:post},
    for $1\leq t\leq T$ and a given set of 
    observations $y_{1:t}$. 
          
     In the sequel, a superscript $i$ means for all $i\in\{1,\dots, N\}$ denoting
     $N$ as the number of particles in an SMC simulation,
     and $a_t^i$ is the $i^{th}$ element of the ancestor vector $A_t$
     where each element in $A$ 
     is the index of the particle at time $t-1$ which is its 
     parent particle at time $t$.
     The method proceeds as described
     in~\cref{alg:smc}; the particular algorithmic choices are detailed in~\cref{ssec:epi_results}. 

    \begin{algorithm}
        \caption{Sequential Mote Carlo}\label{alg:smc}
        \begin{algorithmic}[1]
           
            \State Initialize $\theta_x, \theta_y$ and sample particles $\{x_0^i\}_{i=1}^N\sim \mu(x_0)$
            \State Set $\{w_0^i\}_{i=1}^N = 1/N$.
            \For{$t = 1,\dots,T$}
                \If {resample}
                    \State $a_t^i\sim$ resample $(a_t^{1:N}\given w_{t-1}^{1:N})$
                    \State $w_t^{1:N} \gets 1/N$
                \Else
                    \State $a_t^i\gets i$
                    \State $w_t^{i}\gets w_{t-1}^i/\sum_{j=1}^N\, w_{t-1}^j$
                \EndIf
                \State $x_{t}^{i}\sim p(x_t\given x_{t-1}^{a_{t}^i}; \theta_x)$ // propagate particles
                \If {$Y_t$ is observed}  // update weights
                    \State $w^i_t\gets r(y_t\given x_{t}^{i};\theta_y)\cdot w^i_t / w_{t-1}^{a_{t}^i}$
                \Else
                    \State $w^i_t\gets q(x_t\given x_{t}^{i};\theta_x)\cdot w^i_t / w_{t-1}^{a_{t}^i}$
                \EndIf
        \EndFor
        \end{algorithmic}
    \end{algorithm}
    The choices for reweighting the ensemble on lines 13 and 15 of~\cref{alg:smc} correspond to either 
    the problem of data assimilation or state prediction respectively, depending on the frequency of the observations
    and problem formulation. 
    We define $q(x_t \given x_{t}^{i}; \theta_x)$ to be a Gaussian process and let the observation weighting
    function $r(y_t \given x_{t}^{i}; \theta_y)$
    follow a Gaussian distribution~\cite{del2015sequential,spannaus2022inferring}. In
    this setting, the state transition 
    density $p(x_t \given x_{t-1}^{a_{t}^i}; \theta_x)$
    is implicitly defined through a numerical approximation scheme targeting the 
    system~\cref{eq:SIR}, but other choices are possible as well. The numerical scheme samples from this transition
    density, conditioned on the last time step. Other choices are admissible; see~\cite{chopin2020introduction}
    for options. In the case that $y_t$ is observed, this choice corresponds to the bootstrap filter~\cite{gordon1993novel}.  

    Within the SMC algorithm the resampling step is crucial to its success. Resampling prevents weight degeneracy,
    where a few particles have non-zero weight, 
    and is triggered by examining the effective sample size of the ensemble~\cite{chopin2020introduction,del2006sequential}.
    This step is the primary computational bottleneck within the method as it requires a summation
    across the entire vector $w_{t}^{1:N}$ of weights when resampling is triggered. There have been some
    methods proposed to alleviate this bottleneck as 
    described below~\cite{murray2016parallel,guldas2015practical}. 

\subsubsection{Parallel Prefix Resampling}
\label{sssec:parallel-prefix}

Resampling methods within the SMC literature
scale linearly with the number of particles and require a synchronization step, i.e., gathering
the entire vector of weights;.
These computational considerations have limited the development of 
SMC methods within an HPC framework. 
The computational complexity of popular resampling approaches, such as multinomial, residual,
systematic, or stratified, increases with the number of particles~\cite{murray2016parallel}. 

We use the Thrust~\cite{Thrust} implementation of a parallel prefix-sum algorithm~\cite{sarkka2020temporal} 
which reduces communication between processes~\cite{nicely2019improved},
thus addressing the primary computational bottlenecks inherent to the method.

By incorporating a parallel-scan prefix sum, we reduce the computational time to $\mathcal{O}(\log n)$.
The prefix sum calculation is
a fundamental part of any resampling scheme, as it provides a cumulative
sum of the weights, allowing some particles in low-probability regions to die off, while
propagating those in regions of higher probability. 

In the resampling step, particles are resampled
based on their relative importance in the empirical distribution function.
Running any SMC, or more generally any sequential importance sampling method, 
without a resampling step can lead to weight degeneracy problem, especially in the setting of 
highly-informative observations~\cite{chopin2020introduction}. Weight degeneracy is where
all particles except for one have near-zero weight, leading to a degenerate distribution.
Weight degeneracy
can be avoided by periodically resampling, which is done by eliminating particles with small weights and replacing them with offspring from those
particles that have higher weights, then setting weights for all the particles to be equal.

\section{Algorithmic Design}\label{sec:the_algorithm}

When coupling our ABM simulation with the SMC method we first start from a given set
of initial conditions for the ABM. 
The temporal domain is partitioned into time-steps for the SMC, denoted by $\tau$, and 
steps where information is passed between the ABM and SMC methods, which are given by $t$.
In our setting, $\tau < t$, as there are some number of steps $n$ which comprise each interval $[t, t+1]$.  

We specify the proportion of the 
population which is initially infected at the start of the simulation and
run the ABM forward in time. The proportion of the population which is in
each of the compartments of the epidemiological model at each day are saved as
input to the SMC. Other quantities of interest, such as population proportion 
in the susceptible and recovered compartments, are recorded as well, but are not used 
by the SMC method. 

The SMC sampler then generates a sequence of distributions $\pi_t = \{\pi_\tau\}_{\tau=0}^t$,
where $\pi_t = p(x_t\given y_{1:t})$ and collection of 
samples $\{x_\tau\}_{\tau=0}^t$ from the state space. These samples are frequently the 
proportion of individuals in each compartment of the epidemiological model at time $\tau$.
The method may also be defined over an augmented state space and infer time-varying quantities
such as disease transmissibility or rate of reporting active 
cases~\cite{dureau2013capturing,spannaus2022inferring}.

At time $t$ we compute summary statistics of the distributional approximations obtained from the 
SMC method and pass these summaries back to the ABM. 
As opposed to
setting transition rules \textit{a priori} and \textit{post hoc} model aggregation,
our iterative process
allows for the ABM simulation to better capture the range of possible outcomes.
From these updated distributions, we run the ABM simulation forward in time and repeat
the process.
The pseudo-code for our method is outlined in~\cref{alg:calibration}.

  \begin{algorithm}
        \caption{Combined ABM/SMC calibration framework.}
        \label{alg:calibration}
        \begin{algorithmic}[1]
         \Require $G(V,E)$, $n_i$ (number of infected nodes at $t=0$)
            \State Initialize disease model parameters at $t=0$, $\log\beta_0\sim\mathcal{U}(-20, 1)$ and $\log\gamma_0\sim\mathcal{U}(-20, 1)$.
            \State Randomly pick a set of infected nodes, $|V_i(t=0)|=n_i$
            \State $H(t=0)[V_i] = I$
            \State $H(t=0)[V-V_i] = S$
            \For{$t = 1,\dots,T$}
                \State $V_I(t)$ from agent-based simulation (\cref{alg:abm})
                \State Distribution of $\beta_{t}$, and $\gamma_{t}$ from SMC (\cref{alg:smc}).
        \EndFor
        \end{algorithmic}
    \end{algorithm}

\section{Numerical Experiments}
\label{sec:numerics}

Experiments are conducted on a GPU-enabled Linux workstation with a
NVIDIA RTX A4000 GPU accelerator and CUDA 12.4. The parallel-prefix sum is the Thrust implementation~\cite{Thrust}. Our codebase uses Python 3.11 with JAX~\cite{jax2018github} 
for the Bayesian modeling and is based on~\cite{particles2020github}.
For the agent-based simulation, we used the Repast library to simulate disease 
dynamics for an SIR epidemiological model~\cite{macal2018chisim,collier2020experiences}. 
Single precision 
computations were used throughout our experiments.

\subsection{Epidemiological Modeling Data Assimilation}
\label{ssec:epi_results}

\subsubsection{Epidemiological Model}
The epidemiological model used in our SMC experiments is an SIR compartmental model augmented with
time-varying transmissibility and recovery rates~\cite{del2015sequential,spannaus2022inferring}
presented here in~\cref{eq:SIR}:
\begin{equation}
	\begin{aligned}\label{eq:SIR}
		\der S(t) &= -\beta(t)S(t)\frac{I(t)}{N} \der t, \\
		\der I(t) &= (\beta(t)S(t)\frac{I(t)}{N} - \gamma(t) I(t)) \der t,\\
		\der R(t) &= \gamma(t) I(t)\, \der t,\\
		{\der \log\beta(t)} &= (w_1 - w_2\log\beta(t)) \der t +
		w_3\der B_w(t),\\
        {\der \log\gamma(t)}&= (u_1 - u_2\log\gamma(t)) \der t +
		u_3\der B_u(t),
	\end{aligned}
\end{equation}
where $\{w_i, u_i\}_{i=1}^3$ parameterize the stochastic differential equations,
$\der\log\beta, \der\log\gamma$,
and $B_w(t), B_u(t)$ are their respective Brownian motions.
The differential equations governing the evolution of $\beta$ and $\gamma$
are mean-reverting Ornstein-Uhlenbeck (OU) processes, which allows for some
variance within the model output. We choose this stochastic formulation for 
our experiments to account for the inherent randomness within the ABM simulation and
stochastic nature of a disease outbreak.

On the ABM side, we simulated
an SIR outbreak of a novel pathogen with transmissibility rate $\beta(t)$,
recovery rate $\gamma(t)$, and set the initial infected proportion of the population is 0.2\%.
For each week in the simulation, we run the ABM, tracking the movement of the
population between each compartment within the epidemiological model. The daily counts
for each simulated day are collected and only the number of 
actively infected individuals are sent to the SMC method after one week
of simulated time. Then SMC uses these values as observations for 
an independent simulation of the same week, tracking the proportion of the 
population in each compartment and inferring the 
transmissibility and recovery rates. 
The SMC method computes distributional estimates of the number of 
individuals within each compartment of the epidemiological model and for 
$\beta(t)$ and $\gamma(t)$, the latter two values are passed to the ABM for 
the next week's simulation. This coupling of ABM and SMC obviates
the need for thousands of ABM realizations to quantify the uncertainty therein;
the Bayesian framework of the SMC method readily computes the
necessary estimates.

\begin{figure*}
    \centering
    \includegraphics[width=0.85\textwidth]{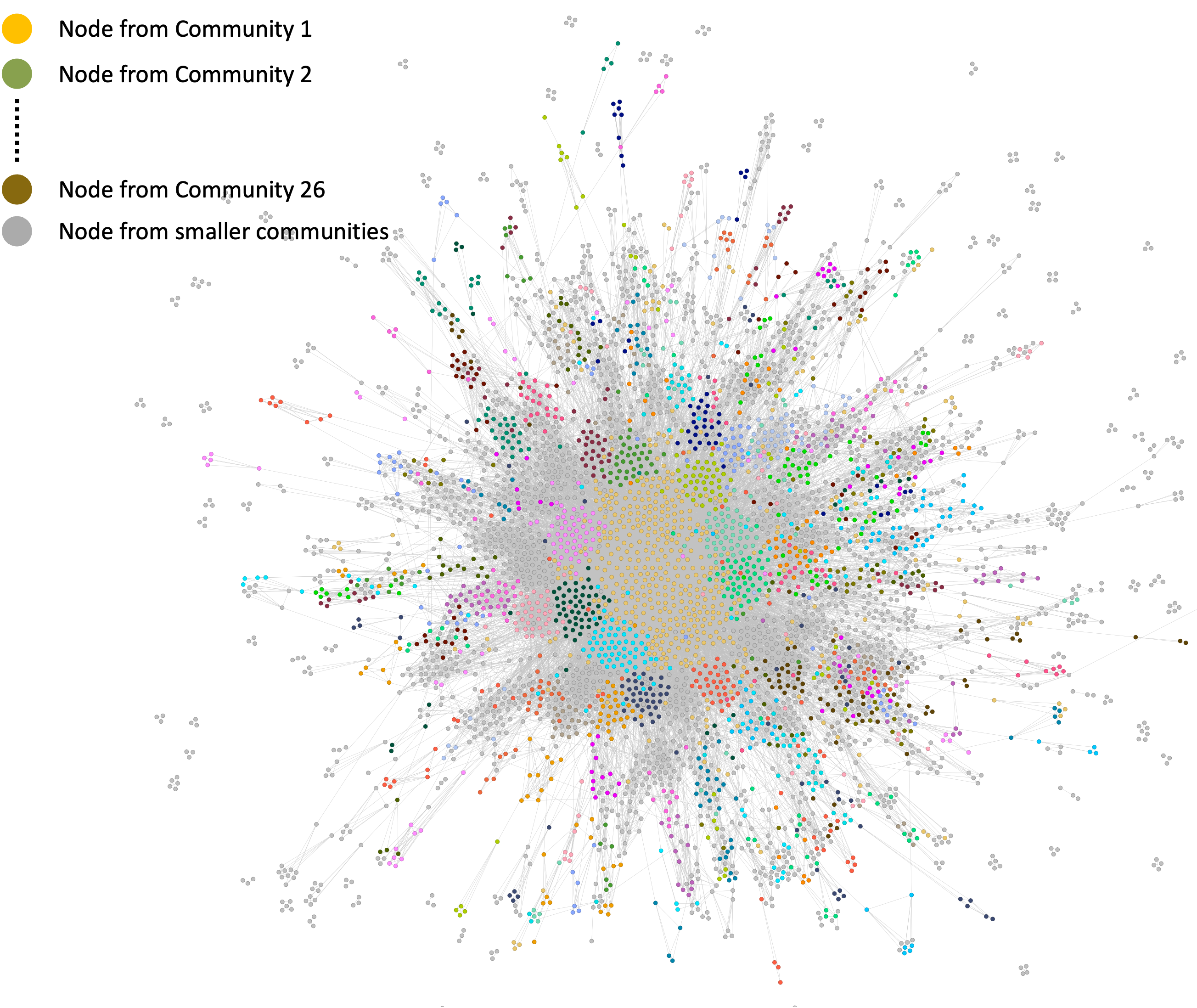}
    \caption{A visualization of the generated BTER network containing 5,000 nodes and 415,100 edges. The top 10\% largest communities are highlighted with different colors. The gray nodes belong to the smaller communities.
         }
    \label{fig:network}
\end{figure*}

\subsubsection{Network-Based Simulation Setup}

We use a synthetic social contact network for the ABM simulation. Real-world direct contact networks are often not available in epidemic modeling due to privacy concerns and a lack of individual-level data. Therefore, 
in our experiments 
we have generated a contact network which emulates the properties found in
real-world networks.

Real-world social contact networks exhibit heavy-tailed degree distribution with community structure~\cite{eubank2006structure}.
Consequently, we employ a realistic synthetic network, a block two–level Erd\"os–R\'enyi (BTER) 
random network.
A BTER network is a well-known graph model that captures observable properties of real-world social networks~\cite{seshadhri2012community}. 
A BTER network generator is scalable and can produce a very large 
network with specific degree distributions and average clustering coefficients. The clustering coefficient of a node $v$ is defined as the probability that neighboring nodes of $v$ in the graph $G$ are also neighbors of each other. A higher average clustering coefficient indicates the presence of well-defined communities or groups in the network.

In this research, we generate a heterogeneous, heavy-tailed BTER network with an average degree of 16.52, consistent with 
previous social network literature~\cite{del2007mixing}. 
Eubank et al.~\cite{eubank2006structure} have conducted a structural analysis of social contact networks derived from detailed traffic studies, reporting a contact network of 1.6 million individuals in Portland, with a clustering coefficient of around 0.57, which is close to our generated network where
the clustering coefficient of the network is approximately 0.55. 
Network structure affects epidemic prevalence; previous literature finds that a heterogeneous network with community structure is less efficient than a homogeneous network in spreading disease and a high average clustering coefficient is a disadvantage for epidemic spread~\cite{liu2003propagation, zanette2002dynamics, wu2008community}. 
\Cref{fig:network} shows a visualization of the generated BTER network for this
study with 5,000 nodes and 415,100 edges.
The dots represent the nodes, and the lines represent the connections between them. The
largest connected component in our BTER network contains more than 92\% of the nodes. We utilize modularity optimization to detect communities in the
network~\cite{blondel2008fast}. It detects 268 communities in the BTER network. The top 10\%
largest communities are highlighted in different colors, with each color representing a
distinct community; the gray nodes belong to smaller communities. 

We leverage the Repast library, an agent-based simulator, for our network-based epidemic simulation~\cite{ozik2021population,collier2022distributed}. Repast is a highly scalable distributed modeling framework that spans multiple processing cores via MPI. 

\subsection{Static Data Assimilation}

In our numerical results, we choose $N=2^{14}$ 
particles in the SMC method, and set $\beta = 0.5, \gamma = 0.1$ for the 
ABM simulation. This choice of $\beta$ and $\gamma$ correspond to a disease with an
$R_0 = \beta/\gamma = 5$, similar to what was observed in the Delta variant of 
SARS-CoV-2~\cite{liu2021reproductive}.
These epidemiological parameter values are hidden to the SMC method and
are only assigned vague uniform priors. The entire simulation is run for 50 days,
with the ABM recording transitions at a daily rate, and passing weekly 
infected counts to the SMC. The 
SMC is a booststrap filter~\cite{chopin2020introduction} and uses a Gaussian process to interpolate 
between observations~\cite{williams2006gaussian}.
Designing bespoke proposals which are more informative than the bootstrap proposal are 
a future direction.

\begin{figure}
    \centering
    \includegraphics[width=\columnwidth]{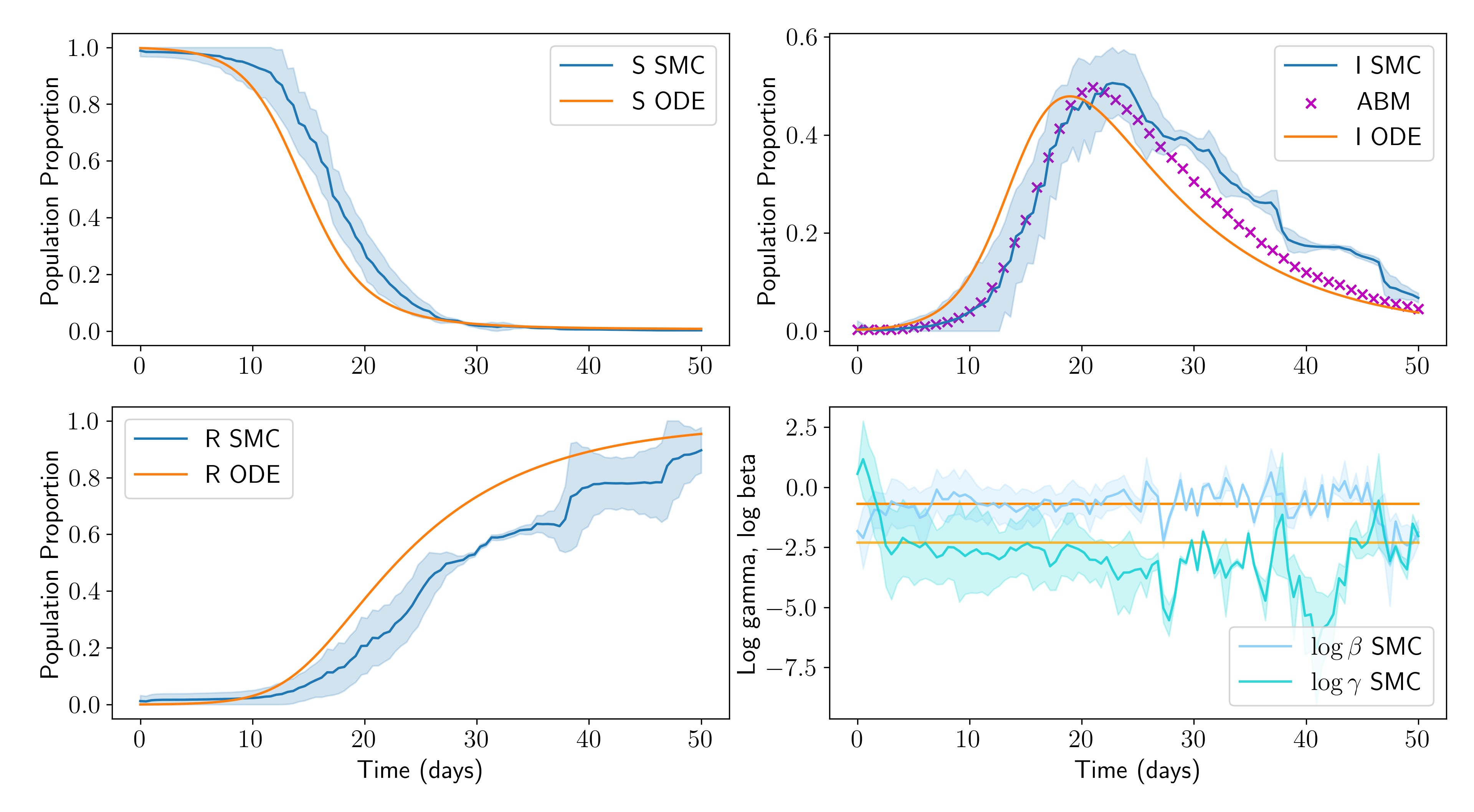}
    \caption{Epidemiological curves from an offline ABM calibration, blue lines
    denote solutions from the SMC, 
    values from the system of epidemiological ordinary differential equations are in orange. }
    \label{fig:static_epi}
\end{figure}

For our first experiment, we consider the setting of offline ABM calibration. We run an 
ABM simulation with a 4,000 node fully connected network. 
This setting allows us to directly compare the solution to the 
system of ordinary differential equations (ODEs) 
against the output from our coupled ABM/SMC method 
to see if we can capture behavior that is unavailable to either the ODE system or
the ABM alone.
To calculate the
ABM transmissibility, we find the ratio of $\beta = 0.5$ to the average node degree and
set $\gamma$ as specified above. We saved the ABM output and used the daily infected counts
as input to our SMC method. 
We interpolated one step between each observation, similar to the
experimental setup in~\cite{del2015sequential} to ensure stability of our solution.
Both started from the same number of infected individuals, but
we did not otherwise inform the SMC method. The initial values
of $\beta$ and $\gamma$ were drawn from vague uniform priors,
specifically $\log\beta_0\sim\mathcal{U}(-20, 1)$, and similarly for 
$\log\gamma_0$. In the SMC, the population 
is normalized, so we chose to model the process and observation  
uncertainty as $\mathcal{N}(0, \sigma^2)$ where $\sigma^2 =0.1$, representing
a 10\%  error on the state estimation and observations.
The results from this simulation are in~\cref{fig:static_epi}.
Our results show that we were able to identify
noisy estimates of the transmissibility and recovery rates used in the ABM. 

As compared with the values from the static 
system of ODEs
we closely followed the ODE curve.
Our method was able to identify the overall trajectory after the
initial uncertainty, reducing the width of the 95/5\% credible intervals as the 
simulation
time progressed. Although our method
 was able to identify  the trajectory of the recovered class,
 we observe greater uncertainity in these estimates, as the credible
intervals are wider here, when compared with those in the susceptible class. 
This is 
due to the uncertainity in our estimates of $\beta$ and $\gamma$, which control 
movement into and out of the infected compartment.

\begin{figure}
    \centering
    \includegraphics[width=\columnwidth]{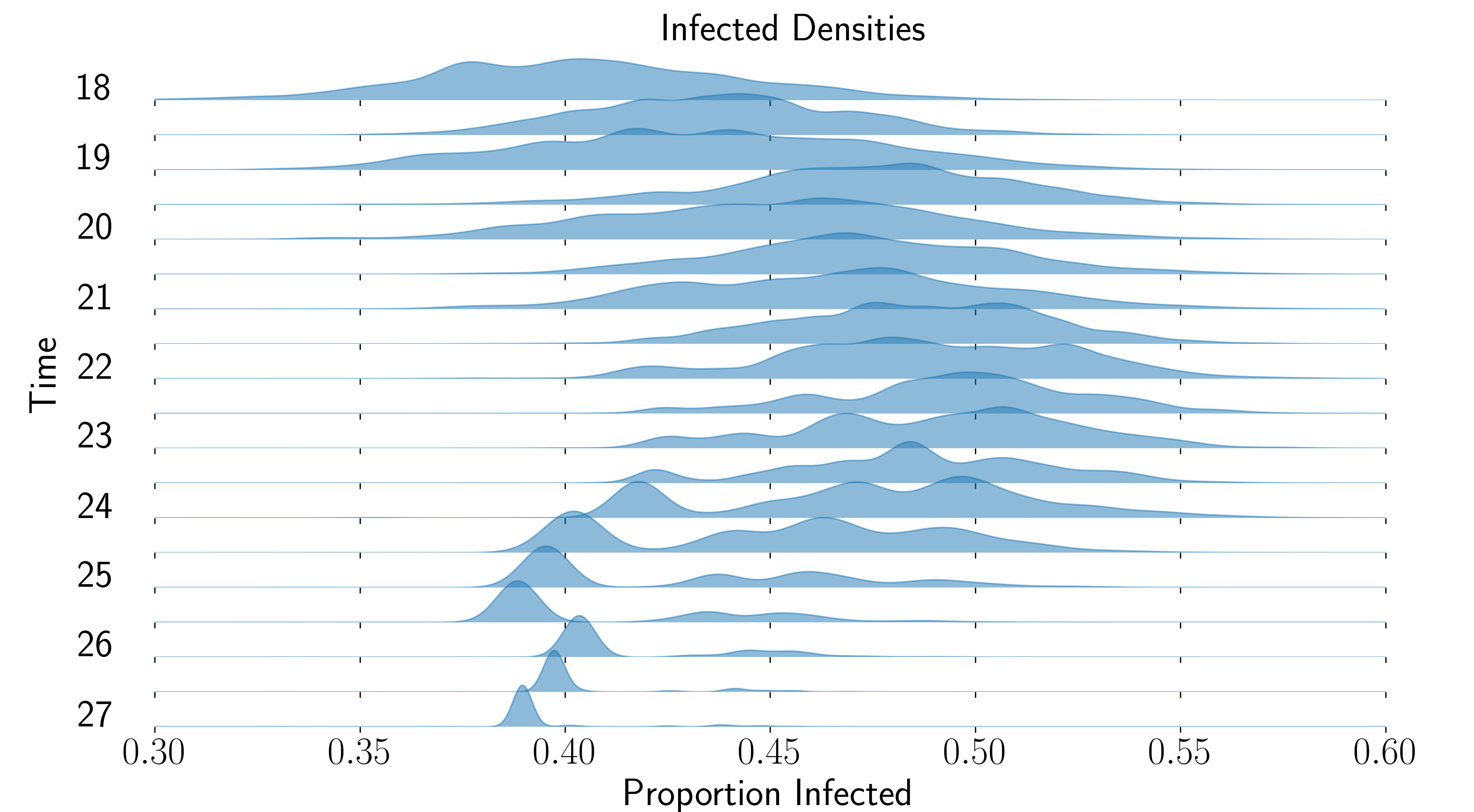}
    \caption{Infected curves from an offline ABM calibration from a subset of the entire time
    window; density curves were estimated through a kernel density estimate. Notice the multi-modality
    of the inferred distributions at the beginning of the time window, which concentrate as the simulation progresses. }
    \label{fig:ridge}
\end{figure}

Turning now to the initial two weeks of the infected proportion,
we observe a trajectory which under-estimates the infected proportion as compared to the ODE solution,
and has wide credible intervals towards the peak of the infected curve, indicating greater uncertainty 
in the distributional estimates at those times. Focusing on a time window starting after the two week mark, 
in~\cref{fig:ridge} we have plotted the kernel density estimates of these distributions to 
understand the source of this greater uncertainty. In this sequence of 
plots, we observe that the distributions start multimodal, with the estimated
density quite diffuse. These densities have a greater spread than those at the 
end of this subset of time, due to the wide credible intervals for $\gamma$. 

As time progresses, we see that these density estimates 
become more concentrated and have a distinct peak by the end of this window. This initial
multi-modalilty is driven by the wide intervals about $\beta$ and $\gamma$ for the 
first half of the simulation window. 
While the value of $\gamma$ decreases for the first few time steps and
identifies the general neighborhood of $\gamma$, we observe wider credible intervals here, as compared with the latter half of the simulation, leading to the 
greater uncertainity in the infected curve for the first half. Additionally, this 
leads to a somewhat later infected peak time, but also more importantly, a
greater proportion of the population being infected. 
This greater infected peak is driven in part by the under-estimation of 
$\gamma$, this extending the recovery period.

This increased variance and large moves in the $\gamma$ curve are reflected in the 
wide credible intervals for the recovered compartment after day 40 in the 
simulation. While both the infected and recovered curve follow the trend
in the ODE solution, the jumps in $\gamma$ are reflected in the 
credible intervals about the recovered proportion due to the structure of the 
differential equations. 

To further decrease the width of the credible intervals for this setting
we may choose different observational densities, or
by calibrating with real data, so that the value of 
$\gamma$ used in the ABM is closer to real-world
conditions. Since the ABM 
does not assume population homogeneity unlike the ODE, increasing the 
variance of the estimated quantities might better capture the dynamics driving recovery.

We observe that the inferred values of the infected curve lag somewhat, but not dramatically so,
behind those from the ODE system,
this is due in part to the delayed infectivity seen in the ABM.
The values of $\gamma$ towards the end of the simulation
directly impacts the number of infected individuals, which
sets up the discrepancy between the inferred and ODE curves. One potential strategy
to mitigate this disagreement, is to use a larger ensemble of particles within the SMC method,
i.e., increase $N$, or another is to increase the variance, $w_3$, within the stochastic differential 
equation, $\der\log\beta(t)$ in~\cref{eq:SIR}.

\begin{figure}
\begin{subfigure}[t]{0.49\columnwidth}
\includegraphics[width=\columnwidth]{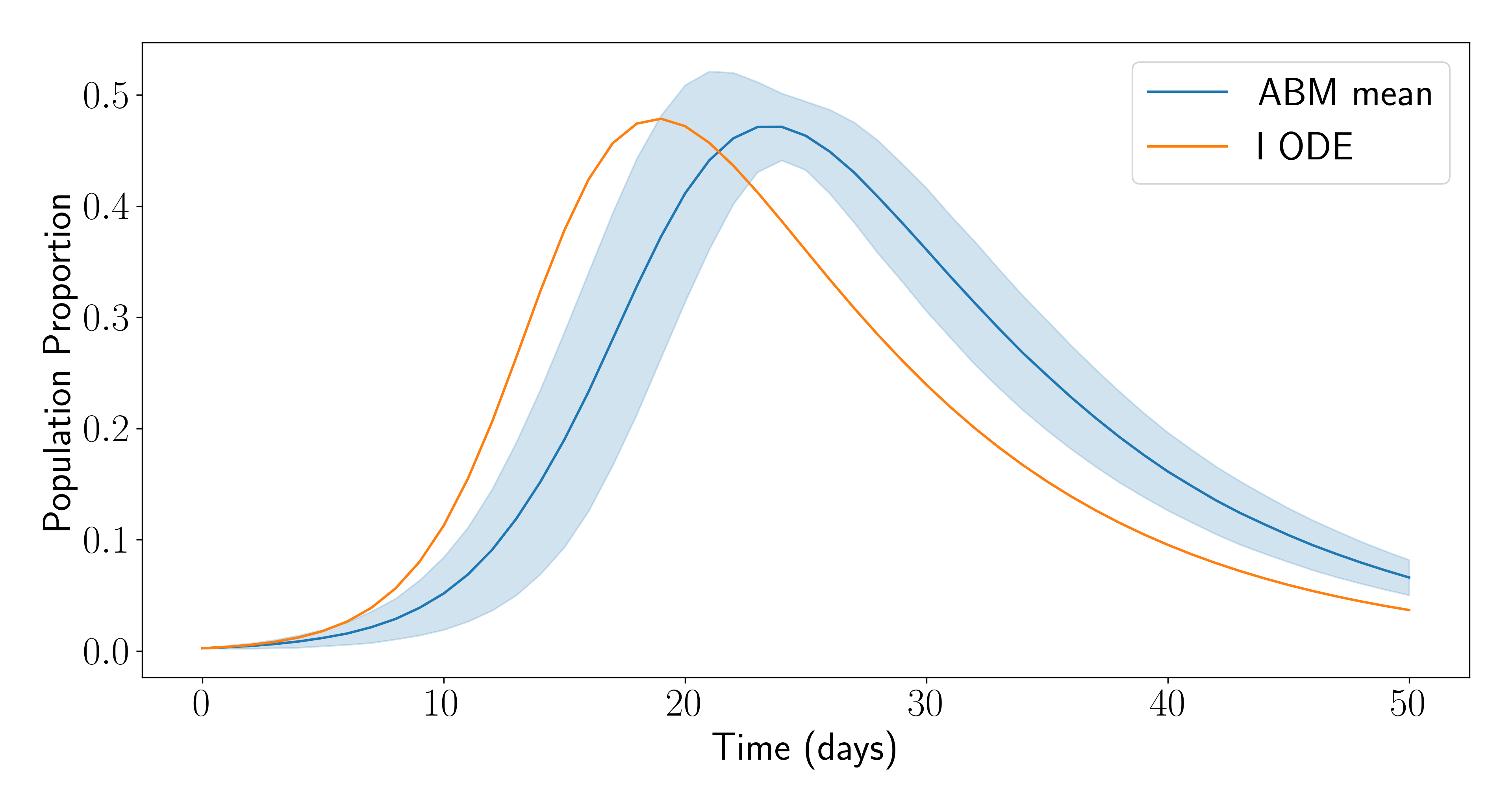}\\
\subcaption{Infected curves from 100 ABM simulations and ODE solution.}
\end{subfigure}%
\begin{subfigure}[t]{0.49\columnwidth}
\includegraphics[width=\columnwidth]{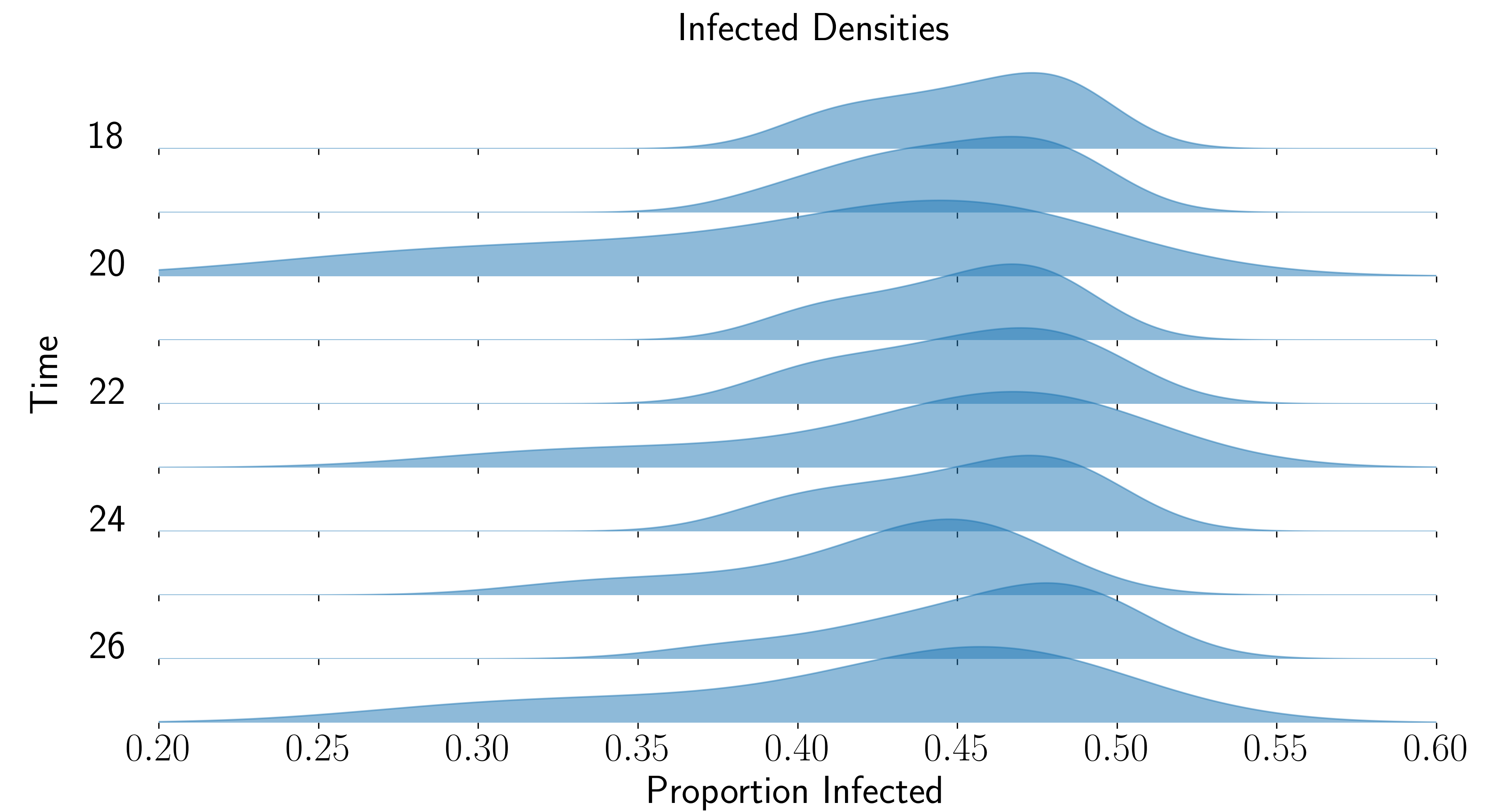}\\
\subcaption{Kernel density estimates from 100 ABM simulations.}
\end{subfigure}%
\caption{In fig (a) we see infected curves from 100 ABM simulations and
their corresponding kernel density estimates in (b). Notice while the 
resulting density estimates are not Gaussian, they do not
exhibit the multi-modality seen in~\cref{fig:ridge}.}
\label{fig:abm_runs}
\end{figure}

Lastly, we examine the inferred densities from 100 ABM simulations
to see if our proposed method can capture behavior which is not
seen by aggregating ABM simulations. In~\cref{fig:abm_runs}
we have the infected curves from 100 ABM simulations 
with the same experimental setup as above and the kernel density estimates (KDEs)
over the same timespan as~\cref{fig:ridge}. Comparing the KDEs between
our coupled method and aggregated simulations, we can see how our model is able
to capture the uncertainity and multimodality present in the system for
this timespan. The aggregated ABM cannot adapt since both the transmissibility
and recovery rates were set \textit{a priori}.
Also note that the infected curve from the ABM lags behind the ODE solution,
as seen in~\cref{fig:static_epi}. Since the ODE system is the 
continuous-time asymptotic limit as the ABM timestep decreases to 0 for 
a fully connected network, the choice of ABM timestep of one day
limits the number of possible interactions that agents may have, 
 causing this delay in infections.

\subsection{Streaming Data Assimilation}

\begin{figure*}
    \centering
    \includegraphics[width=\textwidth]{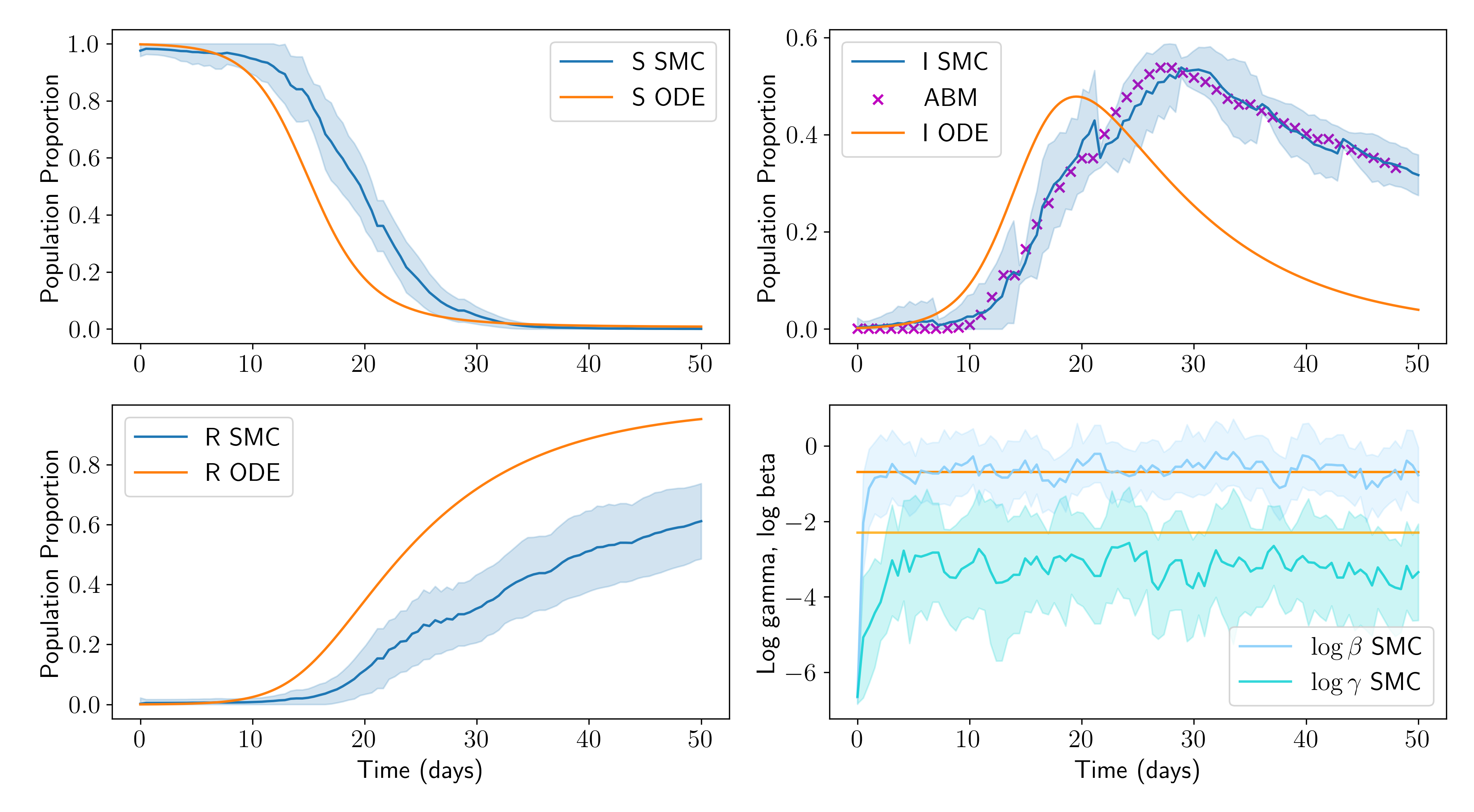}
    \caption{Epidemiological curves from an on-line ABM calibration, blue lines
    denote average solutions from the SMC distributions; 
    values from the system of epidemiological ordinary differential equations are in orange, and ABM values are denoted by purple markers. 
    Shaded regions denote 95\% credible intervals about the mean.}
    \label{fig:dynamic_epi}
\end{figure*}

For our second experiment, we dynamically integrate observational data from our
ABM into our SMC method. Specifically, we randomly initialize the values of 
$\beta$ and $\gamma$ from vague uniform priors for both the ABM and SMC, as in
our previous experiments, and start
both simulations from the same number of infected individuals. After one week 
of ABM simulation time, we pass the number of infected individuals at each day to the 
SMC as observations. We then run the SMC forward for the same duration. At the end of this
interval, we compute distributional estimates of $\beta$ and $\gamma$ and pass these
estimates back to the ABM for the next time window. The epidemiological curves from this 
experiment are presented in~\cref{fig:dynamic_epi}.

For this experiment we compare the results from our coupled ABM and SMC method with 
the solution of the system of ODEs with constant 
values of $\beta$ and $\gamma$, corresponding to~\cref{eq:SIR} the setting where $\der\log(\beta(t)) = \der\log(\gamma(t)) = 0$ for $0\leq t\leq T$.
 When comparing the results in~\cref{fig:dynamic_epi}
from our simulation against the ODE solution we observe generally good agreement 
between the mean inferred value of the susceptible 
class for the duration of the simulation. 
We observe that the recovered class is underestimated
for approximately half of the simulation time,
due to the difference between the ODE solution and 
inferred values for the infected class and the slower rate of recovery, i.e., 
lower value of $\gamma$. 
This discrepancy in the infected curve can be seen from the infected proportion, 
especially for the latter half of the simulation time. 
While the inferred infected curve initially lags behind that from the ODE solution, 
a greater proportion of the population is infected in the coupled ABM/SMC simulation
as compared with the static ODE solution. 
This difference is driven by the value of $\gamma(t)$ which
effecting the infected class in two ways:
1.) it exhibits greater variability than does $\beta(t)$ and
2.) is generally lower than the constant value for the simulation.

When comparing the differences in 
	the infected class for the first half of the simulation window, one 
	can see that the ABM underestimates the proportion of infections, due in part
	to the initial estimate of $\beta$. The initial value significantly underestimated
	the transmissibility of the disease, estimating that for every 10,000 persons
  	fewer than 1 would become infected, when the `true' transmissibility rate
	would yield 5,000 infections. While the inferred value of $\beta$ was 
 able to identify the ODE value within the first few simulation days, 
 the decreased proportion of individuals becoming infected contributed in part
 to the difference between the peak infection times in the coupled ABM/SMC 
 simulation and the ODE solution.

Examining the inferred values for $\beta$ and $\gamma$, we broadly see
two trends: 1.) the inferred $\beta(t)$ was able to recover from the poor initialization and 
identify the constant value used in the ODE solution; and 2.) the inferred
$\gamma$ shows greater variability and wider credible
intervals as compared with $\beta$, both due to the uncertainty in the 
infected class.

The infected class occupies a unique position within our epidemiological
model. From a modeling perspective, it is dependent on each of the other 
compartments and model parameters; from a public health viewpoint, knowing 
its peak value and range of possible outcomes allows public healthy agencies to plan
for various scenarios, assessing hospital capacity and consider if intervention
strategies are needed.  
Examining this curve for the first three simulation weeks, we observe
that follows a similar trend to that of the ODE solution, albeit delayed. 
A greater proportion of the population is infected in the coupled simulation,
due in part to the increased recovery time, i.e., decreased value of $\gamma$, 
and uncertainity in its estimates.

\section{Discussion}
\label{sec:discussion}

We have developed a data assimilation method which augments the state-space
of a partially observed dynamical system
with an ABM. This framework allows for
more realistic epidemiological modeling of a population, as classical ODE methods
assume population homogeneity. Coupling these two approaches yields a tool
capable of quantifying: 1.) the spread of a disease in a heterogeneous population,
and 2.) the possible range of outcomes by assimilating near-real time streaming
data into the model. While the proof-of-concept runs conducted in this study are moderately sized, the SMC-based approach and associated packages like Repast scale well and should be suitable for deployment on HPC systems.

Our results demonstrate the efficacy of this approach
in settings with either static or streaming data. There are a few
directions which would further improve our modeling strategy.
It is widely known within the SMC community that identifying good
proposal distributions, i.e., the pushforward measure of
the ensemble from $t$ to $t+1$, is challenging. This issue is magnified in the 
present setting, where we have specified the functional form of the time-varying
components \textit{a priori} in~\cref{eq:SIR}. The formulation employed here 
assumes some steady-state, where $\log\beta$ and $\log\gamma$ are in some interval
about their mean. This scenario may not always be realistic however. If we consider
the setting of a viral mutation or the emergence of a drug resistant strain, the 
mean-reverting Ornstein–Uhlenbeck processes modeled here would no longer be applicable. Rather, either 
of these scenarios would require formulating a new epidemiological model 
and performing new simulations, thus increasing the time before public health agencies can act. 
One possible solution to this limitation would be to learn the functional form of the dynamics from the data itself, instead of specifying it. This is precisely the goal
of neural differential equations~\cite{chen2018neural,kidger2022neural}.

Another direction which would improve our model's inferences 
is to generate the distributional estimates for the time-varying parameters
from the smoothing distribution as opposed to the filtering distribution
within the SMC model. These two distributions are related~\cite{law2015data},
but provide different information. The filtering distribution provides
$\pi(x_t\given y_t)$ whereas the smoothing distribution considers the entire
observation trajectory $\pi(x_t\given y_{1:t})$, which `smoothes' out the roughness
introduced by the noise in the pushforward measure when going from $t$ to $t+1$. 
This increased path regularity comes at a computational cost, as typical algorithms
scale linearly in time. A recent work~\cite{JMLR:v23:22-0140}
reduces the computational cost to $\mathcal{O}(\log_2 T)$ through a
clever combination step when stitching together the resulting posterior distributions.

\subsection{Conclusions}

Our proposed data assimilation method successfully combines the power of 
a micro-scale epidemiological simulation within a Bayesian framework to quantify
the uncertainty and make model corrections at each time-step. Our data assimilation
framework allows for integrating near-real time data into
an ABM, keeping the results closer to reality as a situation evolves.
Our framework also allows for time-varying parameters to be 
seamlessly included into the underlying epidemiological model. 
We rule out implausible parameter choices within the ABM, thus
reducing model error and providing a range of realistic outcomes.

The resulting epidemiological curves from our static and streaming data experiments
demonstrate that our framework captures the time-varying dynamics of 
a disease outbreak and adapts the model to the dynamics within the data.
By calibrating our ABM with real-world
data, the inferred epidemiological parameters
will be closer to reality than the 
standard ABM modeling setup, as the latter approach cannot dynamically 
react to changes in epidemiological structure. 
Solutions to a system of epidemiological ODEs
also fail to capture the entirety of the system at the necessary fidelity, due to their reliance on
the assumption of population homogeneity. It is through coupling these two strategies within
a Bayesian framework that we may provide a method
to assimilate near-real time data into ABM simulations to
advise public health agencies faced with a novel disease outbreak.

\section*{Acknowledgments}

    This manuscript has been authored by UT-Battelle, LLC, under contract DE-AC05-00OR22725 with the US Department of Energy (DOE). The US government retains and the publisher, by accepting the article for publication, acknowledges that the US government retains a nonexclusive, paid-up, irrevocable, worldwide license to publish or reproduce the published form of this manuscript, or allow others to do so, for US government purposes. DOE will provide public access to these results of federally sponsored research in accordance with the DOE Public Access Plan (http://energy.gov/downloads/doe-public-access-plan).

\section*{Declarations}

\subsection*{Funding}

    This material is based upon work supported by the U.S. Department of Energy,
    Office of Science, Office of Advanced Scientific Computing under Award Number DE-SC-ERKJ422.

\subsection*{Availability of data and materials}
 The code used to create
	the figures and simulation results are  in a git 
	repository:\\ \texttt{https://github.com/aspannaus/abm-uq}.
	
\subsection*{Competing interests}
The authors declare that they have no competing interests.

\bibliographystyle{acm}
\bibliography{refs.bib}

\end{document}